\documentclass{aa}
\usepackage{psfig}
\def\solar {\ifmmode_{\mathord\odot} \else $_{\mathord\odot}$\fi}
\def\Msol {\ifmmode {\,{\it M}\solar} \else $\,M$\solar\fi}     
\def\Rsol {\ifmmode {\,{\it R}\solar} \else $\,R$\solar\fi}     
\def\Lsol {\ifmmode {\,{\it L}\solar} \else $\,L$\solar\fi}     
\newcommand{\less}{\raisebox{-0.6ex}{$\,\stackrel{\raisebox{-.2ex}{$\textstyle<$}}{\sim}\,$}}

\begin{document}
\thesaurus{06     
              (03.20.7;  
               08.12.2;  
               08.12.1;  
               07.19.2)  
    }
\title{The closest extrasolar planet}
\subtitle{ A giant planet around the M4 dwarf Gl~876}
\thanks{Partly based on observations made at the Observatoire de Haute Provence
(CNRS)}
\thanks{Partly based on observations obtained with the swiss 1.2m telescope at the European Southern Observatory
}
\author{X.~Delfosse \inst{1}\inst{2}
   \and T.~Forveille \inst{2}
   \and M.~Mayor \inst{1}       
   \and C.~Perrier \inst{2}     
   \and D.~Naef \inst{1}        
   \and D.~Queloz \inst{3}\inst{1}    
}
\offprints{Xavier Delfosse, e-mail: Xavier.Delfosse@obs.unige.ch}
\institute{   Observatoire de Gen\`eve,
              51 Ch des Maillettes,
              1290 Sauverny,
              Switzerland          
\and
              Observatoire de Grenoble,
              414 rue de la Piscine,
              Domaine Universitaire de S$^{\mathrm t}$ Martin d'H\`eres,
              F-38041 Grenoble,
              France
\and
              Jet Propulsion Laboratory, 
              Mail Stop 306-473, 4800 Oak Grove Drive, Pasadena,
              CA 91109,
              USA
}
\date{Received ; Accepted}
\maketitle
   \begin{abstract}
Precise radial velocity observations of the nearby M4 dwarf Gl~876
with the Observatoire de Haute Provence 1.93~m telescope and the new
1.20~m swiss telescope at la Silla indicate the presence of a jovian
mass companion to this star. The orbital fit to the data gives a period
of 60.8 days, a velocity amplitude of 246m.s$^{-1}$ and an eccentricity
of 0.34. Assuming that Gl~876 has a mass of 0.3{\Msol}, the mass function
implies a mass for the companion of 2/$\sin{i}$ Jupiter masses.
   \end{abstract}
%

\section{Introduction}
The still recent discovery of the first extrasolar planet, around
51~Peg (Mayor \& Queloz, 1995), has since been followed by many
more. The count currently runs to 11 very low mass companions 
(Marcy \& Butler, 1998; Queloz, 1999), with minimum masses
(M~$\sin{i}$) which range between 0.5 and 10 times the mass of
Jupiter. Asides from their Jupiter-like masses, which largely
reflect the sensitivity threshold of current radial velocity
programs,
the known extra-solar planets are a very diverse class. Some of them
have large eccentricities when others have nearly circular orbits, and
their periods range between 3.3 days and 4.4 years. Giant planets can
thus have very much shorter periods than in our solar system, which
clearly does not represent the only possible outcome of planetary
system formation and evolution.

To date on the other hand, planets have mostly been looked for around
solar type stars, and, pulsar companions asides, they have only been
found orbiting such stars. This reflects to some extent an
understandable desire to identify close analogs to our own solar
system, which could perhaps contain life sustaining planets. Also, the
selection function of the radial velocity planet searches has a
relatively sharp optimum around spectral class G. Essentially all
stars hotter than approximately F5 have fast rotation (Wolf et al.,
1982), so that it is impossible to measure their radial velocity to
the $\sim$10m.s$^{-1}$ accuracy needed to detect planets. At the other
end of the mass spectrum, most M dwarfs have slow rotation (Delfosse
et al., 1998a) and their velocity can be measured accurately, as we
discuss below.  Their luminosities however are much lower than those
of solar type stars. At a given distance a much longer integration time is
thus needed to obtain a given radial velocity precision on an M dwarf
than on a G dwarf. All planet search programs have thus understandably
concentrated on solar type stars.

G dwarfs however only represent a small fraction of the disk stellar
population, with the lower mass M dwarfs outnumbering them by about an
order of magnitude (Gliese \& Jahreiss, 1991). It is thus likely that
most planets in our galaxy orbit stars whose mass and luminosity are
significantly lower than the Sun's (Boss, 1995), unless some as yet
unidentified physical process restricts planet formation to the
environment of sufficiently massive stars. It is clearly important
to establish whether such a mechanism exists.

For the last three years, we have been monitoring the radial
velocities of a nominally volume limited sample of 125 nearby M
dwarfs. The two main goals of this large observing program
($\sim$30~nights/year) are to establish the controversial
(e.g. Kroupa, 1995, and Reid \& Gizis, 1997, for two contrasted views)
multiplicity statistics of field M dwarf systems, and to pin down the
still uncertain mass-luminosity relation at the bottom of the main
sequence.  Delfosse et al (1998b) present preliminary results for the
stellar companion search, with 12 new components found in these nearby
M dwarf systems, including the third detached M dwarf eclipsing binary
(Delfosse et al., 1998c).
%
A byproduct of this program, related to the angular momentum dissipation 
of very low-mass stars, is described in Delfosse et al (1998a).

Even though this was not the main focus of the program, we also
realised from the start that for most of these stars we obtain radial
velocity precisions which are sufficient to detect giant planets, if
any exists around them.  We present in this letter the first such
detection, around Gl~876 (BD$-15\degr$6290, LHS~530, Ross~780,
HIP~113020), a V=10.2 M4 dwarf (Reid et al., 1995) at
d~=~4.702$\pm$0.046~pc (ESA, 1997).

Delfosse et al. (1998a) present in detail the observed sample, while
Delfosse et al. (1998b) discuss the observing and analysis technique
at length. We therefore only briefly summarize this information in
section 2. We then proceed to discuss in section 3 the radial velocity
detection of the planetary companion of GL~876.  In section 4 we
consider the implications of this detection and suggest some possible
follow-up observations.


\section{Observing program}
The sample contains the 127 M dwarfs listed in the third edition of
the nearby star catalog (CNS3 preliminary version, Gliese \& Jahreiss,
1991) with a distance closer than 9~pc, a B1950.0 declination above
-16~degrees, brighter than V=15, and without a close much brighter
primary. Observations have been carried out since September 1995 with
the ELODIE fiber-fed spectrograph (Baranne et al., 1996) on the 1.93m
telescope at Observatoire de Haute Provence (OHP). The R=42000 spectra
are wavelength calibrated through simultaneous observations of a
thorium lamp. Since June 1998 some southern stars have
also been observed with the nearly identical CORALIE spectrograph on
the recently commissioned swiss 1.2m telescope at la Silla
(Chile). CORALIE mostly differs from the older ELODIE instrument by
its spectral resolution of R=50000, better sampling
of the spectrograph PSF by the CCD camera pixels, and significantly
better temperature control. The first indications are that these
modifications together result in a substantially improved intrinsic
stability (Queloz et al., in preparation).

The extracted M dwarf spectra are analysed through cross-correlation
with a binary (0/1) template constructed from an observed ELODIE
spectrum of Barnard's star, Gl699 (Delfosse et al., 1998b).  For
slowly rotating stars the resulting velocities have internal standard
errors (photon noise plus low level uncalibrated instrumental
instabilities) which typically range from 10-15~m.s$^{-1}$ for bright M
dwarfs (V{\less}10) to $\sim$75~m.s$^{-1}$ at the magnitude limit of
the sample.  For Gl~876 (V=10.2) typical standard errors are 10 to
20~m.s$^{-1}$, depending on airmass and seeing conditions.  Magnetic
activity is common in M type dwarfs, and may further degrade the
measurement accuracy (Saar et al., 1998). This potential error source
is still uncompletely characterised for M dwarfs, but for slowly
rotating stars ($V\,sin{i}<3km.s^{-1}$) we can already bound it to
$\sigma_{\mathrm Vr}${\less}20m.s$^{-1}$ for our cross-correlation
analysis with the M4 binary template. Within the brighter two thirds
of the sample, a conservative assumption at the present time is 
that we will detect all variables with semi-amplitudes larger than
40m.s$^{-1}$.  Assuming for illustration a 5 years period, this
corresponds to a 1 Jupiter mass (M$_J$) planet orbiting a 0.25~{\Msol}
M4V primary (at 1.8~AU), or to a 2~M$_J$ planet orbiting a 0.6~{\Msol}
M0V primary (at 2.5~AU).

\section{A planet around Gl~876}
Since planet detection was not initially emphasized in the observing
program, its sampling strategy is not optimal for detection of low
amplitude variations on timescales shorter than a few years.  Gl 876
was observed once at each observing seasons in 1995 and 1996 and its
velocity variations became apparent from the three observations obtained 
in late 1997. It was then marked in our program lists as a variable.
This low declination source however became unobservable
from OHP before we could gather more data and determine its orbit.
The commissioning of the swiss 1.2m telescope at la Silla and its
CORALIE spectrograph in June 1998 provided the first opportunity to
obtain 3 additional measurements of this southerly source, which
allowed to finally determine its orbit. These observations were
obtained within two weeks of the first light of this telescope,
providing an encouraging indication on its potential for planet
discovery. An end of night measurement from OHP at a large airmass
provided a confirmation on June 22, just in time to confidently
announce the discovery at the IAU ``Precise Stellar Radial
Velocities'' conference (Victoria, Canada, June 21$^{st}$ to
26$^{th}$). At this conference we learned from G.~Marcy that his group
independently identified the orbit of Gl~876, with orbital elements
compatible with our own determination. Weather permitting, we have
since then attempted to observe Gl~876 at most every three nights, and
often every night. The 1998 data therefore dominate the orbital
solution.

\begin{table}
  \center
  \tabcolsep 0.1cm
  \caption{Orbital elements of GL~876. 
           }
  \begin{tabular}{lccl}
Element  & Value & St. Err. & Unit\\[10pt]
P        &    60.97  &   .19 & Days\\
T       & 2450661.7 &  1.5 & Julian Days \\ 
$e$      &      .336 &   0.019 & \\
$\omega$ &    5.2 &  4.8   & $\deg$ \\
K$_1$       &     248.0 &   6.6  & m.s$^{-1}$\\
V$_0$       &    -1.902 &    0.006   & km.s$^{-1}$\\ 
a$_{1}\sin{i}$       &   0.00131 &       & AU \\ 
f(m)       &   7.8$10^{-8}$  &       & \Msol \\ 
O-C(CORALIE)       &    23 & (rms)      & m.s$^{-1}$ \\
O-C(ELODIE)       &     16 &  (rms)     & m.s$^{-1}$ \\
  \end{tabular}
\raggedright
\noindent
\label{elements}
\end{table}

   \begin{figure}

\psfig{height=5cm,file=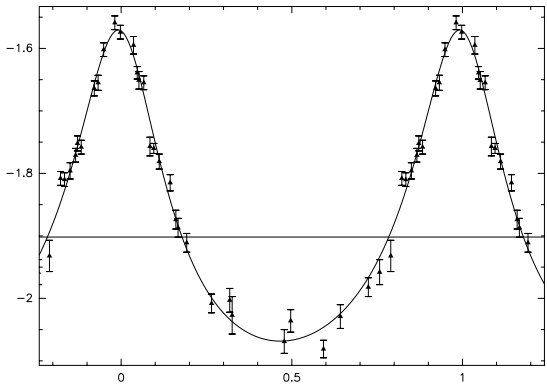,angle=-90}
     \caption{Combined ELODIE and CORALIE radial velocities for Gl~876.
              The solid line is the radial velocity curve for the orbital
              solution.
             }
         \label{rvorbit}
    \end{figure}
The orbital solution is given in Table~1. Preliminary
solutions included a velocity zero point offset between the northern
(ELODIE) and southern (CORALIE) datasets as a free parameter. The two
systems were found to be entirely consistent, and this parameter was
thus held fixed to zero for the final solution. Figure~1
shows the individual radial velocity measurements as a function of
orbital phase (the 16 orbital periods elapsed since the first
measurement make unpractical a display as a function of time; we
however have essentially continuous coverage of one period in June and
July 1998, excluding any possible spectral alias). The orbital period
is two months and the velocity semi-amplitude is $\sim$250~m.s$^{-1}$,
over 10 times the standard error of one radial velocity measurement.
The radial velocity curve implies a moderate but highly significant
eccentricity of $e$=0.34.  

The large amplitude and moderate period of the radial velocity
variation argue strongly for orbital motion as its cause. An
integration of the radial velocity curve implies a minimum physical
motion of $\sim$0.5~\Rsol. This variation is about twice the radius of
an M4 dwarf (Baraffe \& Chabrier, 1996; Chabrier \& Baraffe, 1997),
excluding pulsation as a possible explanation. Gl~876 in addition only
has low level photometric variability, and indeed happens to be one of
the standard stars of the original UBV system (Johnson \& Harris,
1954). It is actually variable, but with low rms amplitudes of 13~mmag
at V, 9~mmag at R and 6~mmag at I (Weis, 1994). The photometric
variations don't phase well at 61~days and appear consistent with a BY
Dra type variability (Sasselov and Cody, private communication).
Since Gl~876 is a very slow rotator (V$\sin{i}<$2km.s$^{-1}$, Delfosse
et al., 1998a), rotational modulation of such low level surface
inhomogeneities cannot explain its large radial velocity
variations. Densely sampled photometry would on the other hand be of
considerable interest to establish the stellar rotation period.

The mass of Gl~876 unfortunately contributes some uncertainty to the
minimum mass of its companion. As a consequence of H$_2$ recombination
in the photosphere and the deepening convection for lower mass stars
(Kroupa et al. 1990), the luminosity does not drop nearly as quickly
per unit mass for mid-M dwarfs as it does for both higher and lower
mass stars (Henry \& Mc~Carthy, 1993), and it has a stronger
metallicity dependence. Between $\sim$0.50{\Msol} and
$\sim$0.18{\Msol}, Mass-Luminosity relations therefore have both
shallow slopes and large intrinsic dispersions (Henry \& Mc~Carthy,
1993). Gl~876 thus belongs to a spectral type range where the mass of
a single star is poorly constrained by its observable characteristics.
Taking at face value either the solar neighborhood observational
mass-luminosity relation of Henry \& Mc~Carthy (1993) or the solar
metallicity models of Baraffe et al. (1998), the absolute magnitudes
of Gl~876 (M$_V$=11.81, M$_J$=7.56, M$_H$=6.96, M$_K$=6.70, Leggett
(1992) and ESA (1997)) imply a mass of 0.30$\pm$0.05{\Msol} for
Gl~876. As an illustration of possible uncertainties however, Delfosse
et al (1998c) measure M~=~0.432 $\pm$0.001{\Msol} and
M$_V$~=~11.7$\pm$0.2 (M$_V$=11.81 for Gl~876) for the brighter
star in GJ~2069A, an M3.5V eclipsing binary which is probably
super-metal-rich ([M/H]$\sim$+0.5).
From its position in colour-colour diagrams (Leggett, 1992), and
from the relative depth of its cross-correlation dips with several
binary templates (Delfosse et al., in preparation), Gl~876 is probably
more metallic than the sun, though not as much as GJ~2069A. 
We adopt a mass of 0.3{\Msol} for the rest of the discussion but warn
that it is uncertain by perhaps 30\%. The minimum semi-major axis and
planetary mass which result from the orbital solution are then
a$\sin{i}$~=~0.20~AU and M$_{2}\sin{i}$~=~2.0M$_J$. They respectively
scale as M$^{\frac{1}{3}}$ and M$^{\frac{2}{3}}$.


\section{Discussion}
The orbital elements of Gliese 876b are worth noting.  In
spite of the low mass of Gl~876, the ice-condensation radius at the
time of planet formation around this star is $\sim$4~AU, only 20\%
lower than around a solar type star (Boss 1995). Once again the
orbital semi-major axis (a~=~0.20 AU) is thus much smaller than
the expected minimum radius for giant planet formation, and some
orbital migration must have occcured. However the observed orbital
separation is also 4 times larger than the measured semi-major axes of
51~Peg, $\tau$~Boo and $\upsilon$~And (0.04-0.05~AU).  The excess of
planets with such small semi-major axes is believed to result from
outward torques which counteract at short distances the inward torque
induced by the interaction of the planet and the protoplanetary disk
(Lin et al. 1996, Trilling et al. 1998). These torques only become
effective at separations significantly smaller than 0.20~AU, and
cannot have played a signficant role for the Gl~876 system.  It
is also interesting to note that the orbit of GL~876b is eccentric
($e$=0.35), while interaction with an accretion disk is expected to
damp any significant orbital eccentricity of a planet (Goldreich and
Tremaine, 1980). Several mechanisms may be called in to explain the
large orbital eccentricities of giant extrasolar planets, but in the
present case the most interesting possibility is probably the chaotic
interaction of several giant planets (Weidenschilling \& Marzari,
1996; Rasio \& Ford, 1996; Lin \& Ida, 1997).  A frequent final result
of such a strong gravitational interaction is a planetary system with
a single giant planet at a moderate semi-major axis, in an eccentric
orbit.  This could thus simultaneously explain the semi-major axis and
the eccentricity.


Contrary to all previously confirmed planets around main sequence
stars, Gl~876b orbits a star which is very different from our Sun,
showing that planetary systems form around stars of widely different
types.  Gl~876 is much less massive than the Sun, $\sim$0.3{\Msol},
and at most only $\sim$150 times more massive than its planet. Its
radius is only three times as large: the radius of Gl~876 is
$\sim$0.3\Rsol (Chabrier \& Baraffe, 1997), while that of Jupiter is
0.10 solar radii.  Gl 876 is also much cooler than the Sun, and much
less luminous. From the observed I-K and V-I colours (Leggett, 1992)
its effective temperature is 3100 to 3250~K (Leggett et al., 1996), 
compared with 6000~K for the solar surface. From its absolute V magnitude
and the  bolometric correction of Delfosse et al. (1998a) it bolometric
magnitude is 9.42, corresponding to 1.35~$10^{-2}$\Lsol. Even though the
planet of Gl 876 is twice closer to its star than Mercury is to
the Sun, the stellar flux at Gl~876b is only ten times the solar 
flux at Jupiter, and lower than the solar flux at Mars. The apropriate 
albedo for Gl~876b is unclear, and, by analogy with Jupiter (e.g. Podolak 
et al., 1993), its thermal balance of Gl~876b may also include a 
substantial contribution from an internal heat source. A detailed 
evaluation of its effective temperature is thus beyond the scope of 
the present letter, but Gl~876b is clearly much too cold to possibly 
sustain liquid water above the 1~bar level.

Gl~876 is closer to us than all other stars orbited by known
extra-solar planets, by at least a factor of 3.  At d=4.7~pc, Gl~876
is the 40$^{th}$ closest stellar system to our Sun, and the 53$^{rd}$
closest star. Since M dwarfs make up $\sim$80\% of the solar
neighbourhood population (Gliese \& Jahreiss, 1991), it is only
natural that the first member of this numerous class found orbited by
a planet is a very nearby one, unless planetary formation would have
selected against low mass stars. 
This discovery weakens such an hypothesis 
but improved statistics would obviously be needed for
a reliable conclusion.

This detection represents an opportunity to confirm a radial velocity
detected planet through astrometry and determine its actual mass, or
at the very least to set a lower limit which is firmly planetary.  Gl
876 is both at least 3 times closer to us than any other star with a
detected planetary companion, and about 4 times less massive (only
0.3{\Msol} instead of about $\sim$ 1{\Msol} for all previous
detections). Despite its relatively short orbital period of 61 days,
the astrometric reflex motion induced by its $\sim$2$M_J$ companion is
therefore unusually large by extrasolar planet standards, with a
minimum semi-major axis of 0.27~milliarcsecond for an edge-on orbit,
and larger by 1/sin(i) for more face-on geometries.  The best single
measurement precision of an astrometric observations is at present
1~milliarcsecond, with the FGS instrument on HST.  A detection is
clearly an ambitious measurement at this time if the orbit is seen
edge-on, and it would need a very determined effort.  A lower limit on
the inclination of $|\sin{i}|>0.25$ on the other hand will be easily
obtained, and would already imply M$_2<$8~M$_J$. In addition, these
observations can be accomplished over the short timescale of one
orbital period, only 2 months.

Finally, it is interesting to note that the measurements of Gl~876
obtained with the new CORALIE spectrometer on the 1.2~m telescope have
residual O-Cs as small as 22~m/s, for a V magnitude of 10.2. This
discovery of a giant planet around a rather faint M4 dwarf illustrates
that this small telescope will powerfully contribute to the search for
extrasolar planets. The application of the cross-correlation technique
to the full wavelength domain (300~nm) of CORALIE compensates to some
extent the disadvantage of the small telescope aperture (Baranne et
al. 1996), and the nearly full-time availability of the telescope for
planet searches will make future period identifications much easier.


\begin{acknowledgements}
  We thank the technical staff and telescope operators of OHP for
  their support during these long-term observations. We thank Claudio
  Melo for obtaining additional measurements for us with CORALIE, and
  the technical teams of the Geneva and Haute Provence Observatories
  for building this excellent instrument, in particular Luc Weber for his
  hard and efficient work on the CORALIE software.  We are grateful to the
  referee, Gilles Chabrier, for his constructive comments, and to Dimitar
  Sasselov for pointing out to us the low level photometric variability of 
  Gl~876. X.D. acknowledges support by the french Minist\`ere des Affaires 
  Etrang\`eres through a ``Lavoisier'' grant.
\end{acknowledgements}


\begin{thebibliography}{}


   \bibitem[1996]{baraffe96} Baraffe I., Chabrier G., ApJ 461, L51.

   \bibitem[1998]{baraffe98} Baraffe I., Chabrier G., Allard F., Hauschild P.,
     1998, A\&A, in press.

   \bibitem[1997]{chabrier} Chabrier G., Baraffe I., 1997, A\&A 327, 1039.

   \bibitem[1995]{boss} Boss A.P., 1995, Science 267, 360.

   \bibitem[1998]{delfosse98a} Delfosse X., Forveille T., Perrier C., 
      Mayor M., 1998a, A\&A 331, 581.

   \bibitem[1998]{delfosse98b} Delfosse X., Forveille T., Beuzit J.L.,
      Udry S., Mayor M., Perrier C., 1998b, submitted to A\&A.

   \bibitem[1998]{delfosse98c} Delfosse X., Forveille T., 
      Mayor M., Burnet~M., Perrier C., 1998c, submitted to A\&A. 

   \bibitem[1997]{ESA} ESA, 1997, The Hipparcos Catalog, ESA SP-1200.


   \bibitem[1980]{goldreich} Goldreich P., Tremaine S., 1980, ApJ 241,425.

   \bibitem[1993]{henry} Henry T.J., Mc Carthy D.W., 1993, AJ~106, 773.

   \bibitem[1954]{johnson} Johnson H.L., Harris D.L., 1954, ApJ~120, 193.

   \bibitem[1990]{kroupaetal} Kroupa P., Tout C.A., Gilmore G., 1990, 
    MNRAS 244, 76.

   \bibitem[1995]{kroupa} Kroupa P., 1995, ApJ~453, 358.

   \bibitem[1991]{gliese} Gliese W., Jahreiss H., 1991, Preliminary
     Version of the Third Catalog of Nearby Stars, as available at
     CDS Strasbourg.
 
 
   \bibitem[1992]{leggett92} Leggett S.K., 1992, ApJSupp.~82, 351.

   \bibitem[1996]{leggett96} Leggett S.K., Allard F., Berriman G., 
     Dahn C.C., Hauschild P., 1992, ApJSupp.~82, 351.

   \bibitem[1996]{lin} Lin D.N.C., Bodenheimer P., Richardson D.C.,
       1996, Nature 360, 606.
 
   \bibitem[1997]{lin97} Lin D.N.C., Ida S., 1997, AJ~477, 781.
 

   \bibitem[1995]{mayor95} Mayor M., Queloz D., 1995, Nature~378, 355.

   \bibitem[1998]{marcy98} Marcy G.W., Butler R.P., ARAA ~36, 57.



   \bibitem[1993]{podolak} Podolak M., Hubbard W.B., Pollack J.B. 1993, 
    in Protostars and Planets III, ed. E. H. Levy, \& J. I. Lunine (Tucson:
      Univ. Arizona Press), 1109.

   \bibitem[1999]{queloz99} Queloz D., 1999, "The new planetary
    systems", in ``Planets outside the solar system'', NATO Advanced
    Study Institute school held at Carg\`ese, France, J. M. Mariotti
    (ed.), in press

   \bibitem[1996]{rasio} Rasio F.A., Ford E.B., 1996, Science~274, 954.
  
   \bibitem[1995]{reid} Reid I. N., Hawley S. L., Gizis J. E., 1995, 
    AJ~110, 1838.
  
   \bibitem[1995]{reid95a} Reid I. N., Gizis J. E., 1997, AJ~113, 2246.

   \bibitem[1998]{saar} Saar S.H., Butler R.P., Marcy G.W., 1998, ApJ~498, 
    L153.

   \bibitem[1998]{trilling} Trilling D.E., Benz W., Guillot T., Lunine J.I.,
    Hubbard W.B., Burrows A., 1998 ApJ~500, 428.

   \bibitem[1996]{weidenschilling} Weidenschilling S.J., Marzari F., 1996,
     Nature 384, 619.

   \bibitem[1982]{wolf} Wolf S.C., Edwards S., Preston G.W, 1982, ApJ~252, 322.

\end{thebibliography}
\end{document}